\documentclass[preprintnumbers,nofootinbib,noshowpacs,eqsecnum,prd]{revtex4}

\usepackage{graphicx,epsfig}
\usepackage{amsmath,amssymb,bbold}
\usepackage{url}

\graphicspath{ {Plots/} }

\makeatletter       

\renewcommand{\p@subsection}{}

\makeatother

\newcommand{\E}{\ensuremath{\epsilon}}

\newcommand{\qb}{\ensuremath{\Bar{q}}}
\newcommand{\qw}{\ensuremath{\tilde{q}}}

\newcommand{\gw}{\ensuremath{\tilde{g}}}

\newcommand{\mqw}{\ensuremath{m_{\tilde{q}}}}
\newcommand{\mgw}{\ensuremath{m_{\tilde{g}}}}

\begin{document}

\title{ \hfill{\footnotesize{DESY 09-030}} \\[-2mm]
        \hfill{\footnotesize{PITHA 09/08}} \\[-2mm]
        \hfill{\footnotesize{PSI-PR-09-02}} \\[6mm]
Gluino Polarization at the LHC
}

\author{M.~Kr\"amer$^{1}$, E.~Popenda$^1$, M.~Spira$^2$, and
  P.~M.~Zerwas$^{3,1}$\\[-3mm]
  \mbox{ } }
\affiliation{ $^1$ Institut f\"ur Theoretische Physik E, RWTH Aachen University, D-52074 Aachen, Germany    \\
  $^2$ Paul Scherrer Institut, CH--5232 Villigen PSI, Switzerland                \\
  $^3$ Deutsches Elektronen-Synchrotron DESY, D-22603 Hamburg, Germany
} \date{\today}

\begin{abstract}
  Gluinos are produced pairwise at the LHC in quark-antiquark and
  gluon-gluon collisions: $q \bar{q},gg \to \gw \gw$. While the
  individual polarization of gluinos vanishes in the limit in which
  the small mass difference between L and R squarks of the first two
  generations is neglected, non-zero spin-spin correlations are
  predicted within gluino pairs.  If the squark/quark charges in
  Majorana gluino decays are tagged, the spin correlations have an
  impact on the energy and angular distributions in reconstructed
  final states. On the other hand, the gluino polarization in single
  gluino production in the supersymmetric Compton process $g q \to \gw
  \qw_{R,L}$ is predicted to be non-zero, and the polarization affects
  the final-state distributions in super-Compton events.
\end{abstract}

\maketitle


\section{Introduction}

\begin{figure}[t]
\epsfig{figure=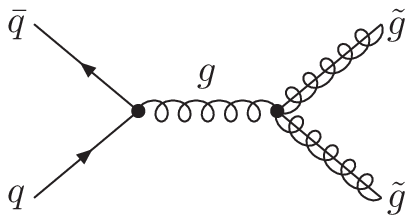, width=3.5cm}
\hspace{0.3cm}
\epsfig{figure=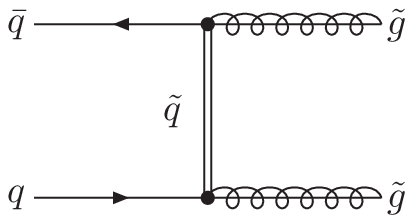, width=3.5cm}
\hspace{1cm}
\epsfig{figure=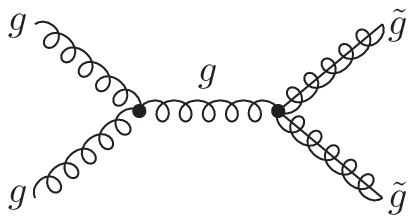, width=3.5cm}
\hspace{0.3cm}
\epsfig{figure=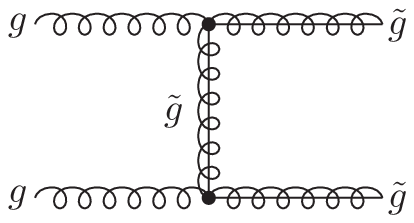, width=3.5cm}
\\[1ex]
\makebox[7.5cm]{(a)}
\hspace{1cm}
\makebox[7.6cm]{(b)}
\\[3ex]
\epsfig{figure=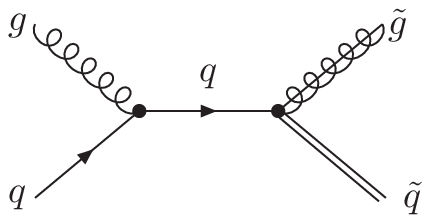, width=3.5cm}
\hspace{0.3cm}
\epsfig{figure=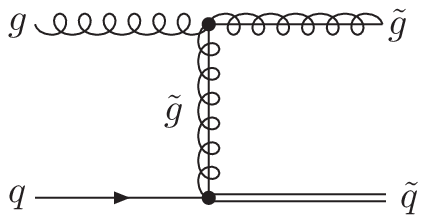, width=3.5cm}
\makebox[8.5cm]{}
\\[1ex]
\makebox[7.5cm]{(c)}
\hspace{1cm}
\makebox[7.6cm]{}
\caption{Feynman diagrams for gluino production in quark annihilation
  (a), gluon fusion (b) and in the super-Compton process (c).}
\label{fig:dia1}
\end{figure}

Gluinos in the Minimal Supersymmetric Standard Model [MSSM]
\cite{susy1,susy2,susy3} can be produced copiously at the LHC, {\it
  cf.} Refs.~\cite{dawson,Beenakker}; gluino pairs in quark-antiquark
and gluon-gluon collisions, Figs.~\ref{fig:dia1}(a/b),
\begin{eqnarray}
   q \bar{q} &\to& \gw \gw 
   \label{eq:qq}\\
   g g       &\to& \gw \gw \,,
   \label{eq:gg}
\end{eqnarray}
and single gluinos in association with squarks in the super-Compton
process, Fig.~\ref{fig:dia1}(c),
\begin{equation}
   q g \to \qw \gw \,.
   \label{eq:qwgw}
\end{equation}
After the gluinos decay, the final-state energy and angular
distributions will in general depend on the degree of polarization
with which the gluinos are generated.\footnote{Spin measurements of
  supersymmetric particles {\it sui generis} at the LHC are widely
  discussed in the literature; for a sample of methods that address
  the impact of spin on differential distributions see
  Ref.~\cite{lhcspin}.}  Thus the distribution of experimentally
reconstructed events is affected by the polarization of the gluinos.

However, polarization effects are to a large extent expunged by the
Majorana character of the particles, in particular when parton charges
are not measured. For example, jet angular distributions
in{\footnote{The indices R,L denote the $\pm$ helicities of quarks and
    antiquarks.}}
\begin{equation}
   \gw \to q_{R,L}\,  \qw^\ast_{R,L} \;\;\; {\rm and} \;\;\;   \gw \to \bar{q}_{L,R}\,  \qw_{R,L}
   \label{eq:qqwb}
\end{equation}
follow, mutually, the $[1 \pm \cos\theta]$ law with regard to the
gluino spin vector so that the sum of the spin-dependent terms
vanishes, as a result of $CP$-invariance if $q,\qw$ charges are not
analyzed. [For the sake of simplicity we will restrict ourselves to
the class of SPS1a/a$'$-type scenarios \cite{SPS} in which the gluinos
are heavier than the squarks so that the complexity of gluino decay
patterns is reduced to the maximum extent possible.]

If gluinos, on the other hand, were Dirac particles $\gw_D$ as may be
formalized in N=2 hybrid models \cite{CDFZ,NT,Fayet,sigma}, the
conservation of the Dirac charge $D$, with $D = +1$ for $\gw_D, \qw_R$
and $-1$ for $\gw_D^c, \qw_L$~\cite{CDFZ}, allows only the production
of specific pairs of supersymmetric particles and antiparticles:
$\gw_D \gw_D^c;\, \gw_D \qw_L,\, \gw_D^c \qw_R$ {\it etc.}, followed
by specific decay patterns: $\gw_D \to q_L \qw_L^\ast$ {\it but not}
$\gw_D \to \bar{q}_R \qw_L$, and crosswise for R-squarks/anti-squarks.
This gives rise to stringent restrictions on polarization and
spin-correlation effects. Observing or not observing such correlations
thus signals the Dirac/Majorana character of gluinos.

\vspace{4mm}

{\bf 1) {\underline{Gluino pairs}}:} The masses of L and R-squarks of
the first two generations are generally close to each other.
Neglecting the small mass differences, the super-QCD action of the
first two generations becomes invariant under $P$-transformation
combined with the exchange of $\qw_L \leftrightarrow \qw_R$.  As a
result, the single polarization of gluinos vanishes in the pair
production processes (\ref{eq:qq}) and (\ref{eq:gg}).  However, spin
correlations within the gluino pair are non-trivial.

\begin{figure}[t]
\epsfig{figure=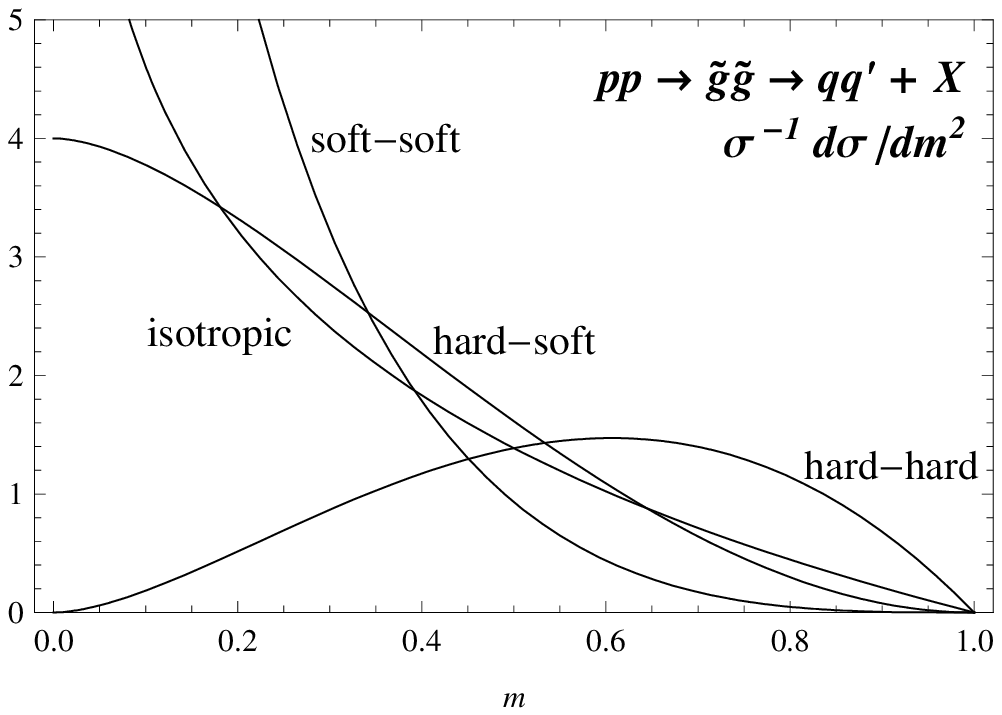, height=5.7cm}
\hspace{0.7cm}
\epsfig{figure=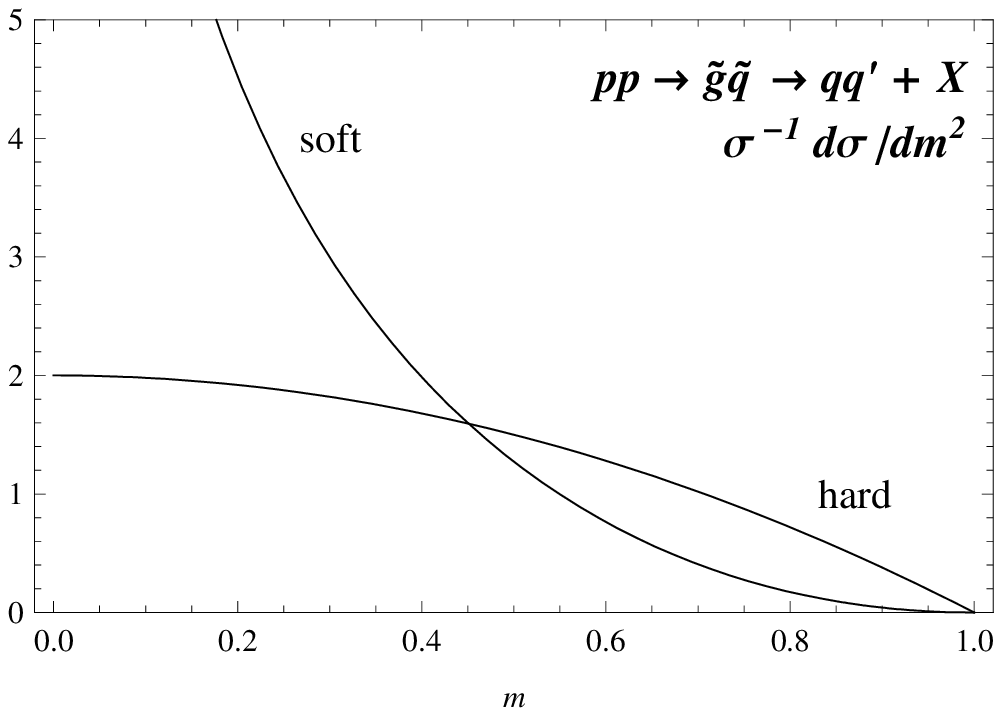, height=5.7cm}
\makebox[8.5cm]{(a)}
\hspace{1cm}
\makebox[7.5cm]{(b)}
\caption{Invariant near-jet jet mass distributions in the gluino pair
  production processes (a) and the super-Compton process (b) for
  various combinations of hard and soft fragmentation $\tilde{g} \to
  q$.}
\label{fig:massdis}
\end{figure}

As argued before, the sum of polarization asymmetries adds up to zero
in $CP$-invariant theories, if the charges are not measured in the
gluino decay modes. Hence, the jets which originate in equal shares
from quarks and antiquarks in (\ref{eq:qqwb}) are isotropically
distributed. The distribution in the scaled jet-jet mass, $m = 2
M_{\rm jj}/\sqrt{s}$ in the decay $\gw\gw \to q q'+X$ is then
moderately soft,
\begin{equation}
   {\rm isotropic:} \;\;\; \sigma^{-1} d\sigma / d m^2 =  \log{m^{-2}}
   \label{eq:isotropic}
\end{equation}
for asymptotic energies $s \gg M^2_{\gw} \gg
M^2_{\qw}$~\cite{Choi:2006mt,lhcspin}.  However, charges can be tagged
in $\tilde{u}_L,\tilde{d}_L$ decays to charginos, $\qw_L \to q
\tilde{\chi}_1^{\pm} \to q l^{\pm}\nu_l\tilde{\chi}_1^0$. This is
useful in scenarios in which $\tilde{\chi}^0_2$ is not a pure wino or
bino state so that the branching ratios of $\tilde{u}_L$ and
$\tilde{d}_L^\ast$ are different. In such scenarios polarization
effects do not completely cancel among the decay chains $\gw \to
\bar{u}_R \tilde{u}_L \to \bar{u}_R d \tilde{\chi}_1^+$ and $\gw \to
d_L \tilde{d}_L^* \to d_L \bar{u} \tilde{\chi}_1^+$, which result in
identical final state charges, but have opposite polarization
signatures.  Tagging of top and bottom charges in the third generation
can be exploited in any case.  In the asymptotic limit in which gluino
fragmentation to quarks [plus accompanying squarks] is either hard or
soft, {\it i.e.}  $\sim 2z $ or $2(1-z)$ for quarks emitted
preferentially parallel or anti-parallel to the gluino spin, various
configurations can be realized for the invariant mass distributions of
the near-jet pairs\footnote{The far-jets will be taken into account
  properly in the detailed phenomenology subsections.} in $\gw\gw \to
q q'+X$ final states \cite{Choi:2006mt}:
\begin{eqnarray}
   {\rm hard-hard:} \;\;\; \sigma^{-1} d\sigma / d m^2 
                           &=& 4 m^2 \log{m^{-2}}                          \nonumber \\
   {\rm hard-soft:} \;\;\; {\phantom{\sigma^{-1} d\sigma / d m^2}} 
                           &=& 4 \left[(1-m^2) - m^2 \log{m^{-2}} \right]  \nonumber \\
   {\rm soft-soft:} \;\;\; {\phantom{\sigma^{-1} d\sigma / d m^2}}
                           &=& 4 \left[(1+m^2)\log{m^{-2}} - 2 (1-m^2) \right]        \,.
\end{eqnarray}
The four mass distributions are compared with each other in
Fig.~\ref{fig:massdis}(a).  Evidently, the gluino polarization leads
to distinct patterns in the invariant jet-jet distributions.

If no charges are measured, the MSSM Majorana theory predicts the
moderately soft mass distribution Eq.~(\ref{eq:isotropic}) of the
near-quark jets in the gluino decays. In a gluino Dirac theory, on the
other hand, pairs $\gw_D \gw_D^c$ are generated decaying to
$\tilde{q}_L \tilde{q}_L^\ast$ {\it but not} $\tilde{q}_L \tilde{q}_L$
final states \cite{CDFZ}. Such a theory would therefore predict an
invariant mass distribution of the hard-hard and soft-soft types for
aligned gluino spins. Thus, the shape of the mass distributions
discriminates between standard Majorana and other Dirac supersymmetric
theories.

\vspace{4mm}

{\bf 2) {\underline{Super-Compton process}}:} Since the processes
Eq.~(\ref{eq:qwgw}) are maximally $P$-violating if L and R-squarks are
discriminated by their decay patterns, the produced gluinos are
polarized.  The polarization can be measured by exploiting the same
methods for the first two and the third generation as outlined above.
The polarization reflects itself in the invariant mass distribution of
the near-quark jet in the gluino cascade with the primary squark decay
jet, for asymptotic energies:
\begin{eqnarray}
   {\rm hard:} \;\;\; \sigma^{-1} d\sigma / d m^2
                      &=& 2\,(1-m^2)  \nonumber \\    
   {\rm soft:} \;\;\; {\phantom{\sigma^{-1} d\sigma / d m^2}}
                      &=& 2\,[\log m^{-2} - (1-m^2)] \,. 
\end{eqnarray}
The two distributions are displayed in Fig.~\ref{fig:massdis}(b),
exhibiting quite different shapes for the hard and soft decay
configurations.

Switching from Majorana to Dirac gluinos does not affect the degree of
gluino polarization. Nevertheless, as a result of $D$-conservation
$\gw_D$ states are generated only in association with $\qw_L$ states,
and $\gw_D^c$ states only in association with $\qw_R$ states, and $cc$
pairs correspondingly \cite{CDFZ}.  The decays of the $\gw_D^{(c)}$
particles follow the pattern discussed earlier.

\medskip

The report includes two central sections. In Section 2 we analyze
spin-spin correlations in gluino pair production, followed by the
discussion of the super-Compton process in Section 3. In both sections
we present analytical results at the parton level, and we illustrate
the spin effects by calculating jet-jet invariant mass distributions
for gluino production and decay at the LHC. While only the basic
theoretical points are elaborated to emphasize the salient features of
the spin correlations, the analysis nevertheless demonstrates the
potential impact of spin-correlations for high-precision supersymmetry
studies at the LHC in the future. Section 4 concludes the report.

\section{GLUINO POLARIZATION IN PAIR PRODUCTION}
\subsection{Gluino Production at the Parton Level}

\noindent
The $s$ and $t$-channel exchange mechanisms, complimented by the
$u$-channel exchanges, are shown in Figs.~\ref{fig:dia1}(a) and (b)
for quark-antiquark and gluon-gluon collisions. For unpolarized beams
the parton-parton cross sections are given by~\cite{dawson,Beenakker}
\begin{eqnarray}
   \sigma[q \bar{q} \to \gw\gw] &=& \frac{\pi \alpha_{s}^2}{s} \beta
                                    \left(\frac{8}{9} + \frac{4 m_{\gw}^2}{9} \right)
                                    - \frac{\pi \alpha_s \hat{\alpha}_s}{s}
                                    \left[ \left( \frac{2 m_{\gw}^2}{3}+\frac{m_{-}^4}{6}\right) L
                                    + \beta \left( \frac{4}{3}+\frac{2 m_{-}^2}{3}\right)\right] \nonumber \\
                                & & + \frac{\pi \hat{\alpha}_s^2}{{s}}
                                    \left[ \left( \frac{16 m_{-}^2}{27}+\frac{4 m_{\gw}^2}{27 \left(2-m_{-}^2 \right)}\right) L
                                    +\beta \left( \frac{32}{27}+\frac{32 m_{-}^4}{27 \left(4\mqw^2
                                    + m_{-}^4\right)}\right) \right]\nonumber
\end{eqnarray}                   															
\begin{eqnarray}
   \sigma[gg \to \gw\gw]        &=& \frac{\pi \alpha_{{s}}^2}{s} \left[ \left( \frac{9}{4} +\frac{9 m_{\gw}^2}{4} -\frac{9 m_{\gw}^4}{16}\right)
                                    \log \left(\frac{1+\beta}{1-\beta} \right)
                                    - \beta \left( 3+\frac{51 m_{\gw}^2}{16} \right)\right]
\end{eqnarray}
where
\begin{eqnarray*}
   L&=&\log \frac{1+{\beta} -{m_{-}^2}/{2}}{1-{\beta} -{m_{-}^2}/{2}}
      \;\;\; {\rm with} \;\;\; \beta=\sqrt{1-m_{\gw}^2} \,,                            \\[1.5mm]
      \alpha_s&=&g_s^2/4\pi \;\;\; {\rm and} \;\;\;
      \hat{\alpha}_s=\hat{g}_s^2/4\pi        \,, 
\end{eqnarray*}
if the polarization of the gluinos in the final state is not measured.
All masses are scaled by the parton beam energy $\sqrt{{s}}/2$ in the
parton-parton c.m. frame, {\it i.e.}
\begin{equation}
m_{\gw,\qw}=2M_{\gw,\qw}/\sqrt{{s}}\;\;\;\; {\rm and} \;\;\;\; 
m_{-}= 2 \sqrt{|M_{\gw}^2-M_{\qw}^2|/s} \,.
\end{equation}
The couplings $g_s$ and ${\hat{g}}_s$ are the QCD gauge and Yukawa
couplings, respectively, which are identical in super-QCD. Though
noted here in Born approximation \cite{dawson}, the cross sections are
known more accurately at next-to-leading order in
super-QCD~\cite{Beenakker} and including threshold
resummations~\cite{Kulesza}. The angular distributions for gluino pair
production read:
\begin{eqnarray}
   \frac{d\sigma}{d\Omega} [q \bar{q} \to \gw\gw] &=& \frac{\beta}{{s}} \left[\frac{\alpha_{s}^2}{6}
                                                      \left(2+m_{\gw}^2-\kappa_{+}\kappa_{-}\right)
                                                      +\frac{2 \alpha_{s} \hat{\alpha}_{s}}{3} \,
                                                      \frac{m_{\gw}^2+\kappa_{+}^2}{m_{-}^2-2\kappa_{+}}\right.
   \label{eq:angdis}\nonumber \\
                                                  & & + \left. \hat{\alpha}_{s}^2 
                                                      \left( 
                                                         \frac{32}{27} \, \frac{\kappa_{+}^2}{(m_{-}^2-2\kappa_{+})^2}
                                                        + \frac{4}{32} \, 
                                                             \frac{m_{\gw}^2}{(m_{-}^2-2\kappa_{+})(m_{-}^2-2\kappa_{-})}
                                                      \right)\right]^{\mathcal{S}}\nonumber\\ [2mm]
   \frac{d\sigma}{d\Omega} [gg \to \gw\gw]        &=& \frac{\alpha_s^2 \beta}{{s}} \, \frac{9}{32} \,
                                                      \left[\frac{\kappa_{+}^2\kappa_{-}^2-2m_{\gw}^2(m_{\gw}^2-1)}
                                                      {2 \kappa_{+}\kappa_{-}}
                                                      +2 \, \frac{\kappa_{+}\kappa_{-}-m_{\gw}^2(m_{\gw}^2-\kappa_{+})}{\kappa_{+}^2}
                                                      -\frac{2\kappa_{+}\kappa_{-}+(\kappa_{+}-\kappa_{-})m_{\gw}^2}{2\kappa_{+}}
                                                      \right]^{\mathcal{S}} \!\! .
\end{eqnarray}
We use the definitions
\begin{eqnarray}
   \kappa_{\pm}&=&1\pm\beta\cos\theta \nonumber\\
   \left[F\right]^{\mathcal{S/A}}&=&F(\cos\theta) \pm F(-\cos\theta)\,.
\end{eqnarray}
from this point on.  The polar angle $\theta$ denotes the gluino
flight direction with respect to the incoming particle momenta in the
parton-parton c.m. frame. The angular distribution is forward-backward
symmetric for the production of Majorana pairs. For the sake of
brevity, the cross sections Eq.~(\ref{eq:angdis}) will be denoted in
the following by ${\mathcal{N}}_{q \bar{q}}$ and ${\mathcal{N}}_{gg}$
for the $q \bar{q}$ and $gg$ channels, respectively.

Quite generally, non-trivial polarization effects are generated in
super-QCD introduced by the parity violating quark-squark-gluino
Yukawa couplings, while spin-spin correlations among the final-state
gluinos are generated also by parity conserving gluon-gluino-gluino
interactions.  However, parity-violating effects in the first two
generations of super-QCD are strongly suppressed by the small
differences between the squark masses. In fact, the super-QCD action
is invariant under the $P$-transformation supplemented by L/R-squark
exchange in the limit of equal-mass L/R squarks. Since the modulus of
the potential gluino polarization vector $|M^2_L - M^2_R|/[M^2_L +
M^2_R] \sim 10^{-2}$ is expected to be small \cite{Eva}, we neglect
these effects in the present analysis.

\begin{figure}[t]
\epsfig{figure=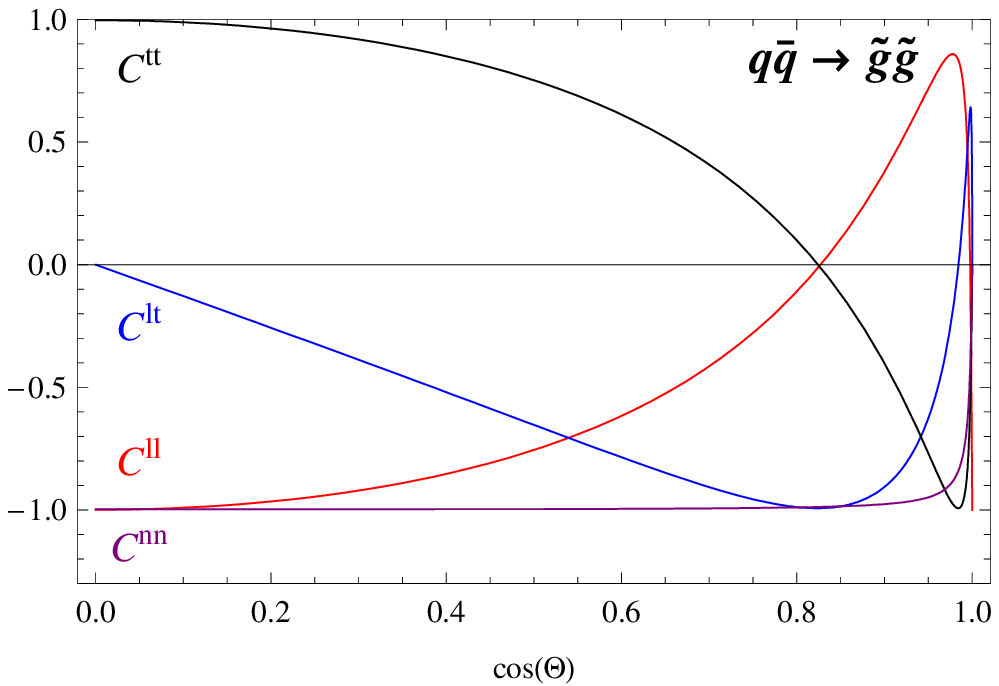, height=5.5cm}
\hspace{0.7cm}
\epsfig{figure=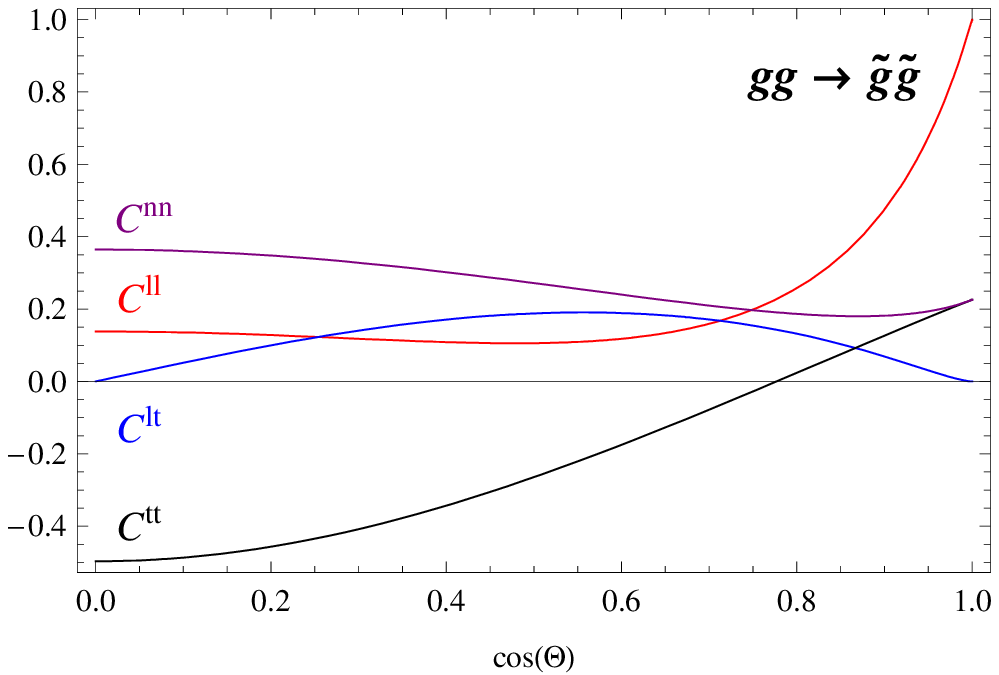, height=5.4cm}
\makebox[9cm]{(a)}
\hspace{1cm}
\makebox[7cm]{(b)}
\caption{Correlation matrix elements for ${\sqrt{{s}}} = 2$ TeV in the
  quark annihilation (a) and gluon fusion (b) channels.}
\label{fig:qqcorr}
\end{figure}

Spin-spin correlations are conveniently described by the tensor
${\mathcal{C}}_{\mu\nu}$ following the formalism developed in
Ref.~\cite{kuhn}:
\begin{equation}
   \frac{d\sigma(s_1,s_2)}{d\Omega} = \frac{d\sigma}{d\Omega}\, 
   \frac{1}{4} \, \left[1 + {\mathcal{C}}_{\mu\nu} s_{1}^{\mu} s_{2}^{\nu} \right] \,.
\end{equation}
The two gluinos in the parton-parton c.m. frame are assigned the spin
vectors $s_1$ and $s_2$; they are related to the spin vectors
${\check{s}}_{1,2}$ in the gluino rest frames by respective Lorentz
transformations $s_{1,2} = \Lambda_{1,2} {\check{s}}_{1,2}$, the
corresponding matrix $\check{\mathcal{C}}$ associated with the general
spin-density matrix, see e.g. \cite{Bern}.  Choosing the
$\hat{z}$-axis along one of the gluino flight directions, the
$\hat{x}$-axis transverse to this vector within the production plane
and pointing into the obtuse wedge between initial and final-state
momenta, and the $\hat{y}$-axis normal to the production plane, the
longitudinal, transverse and normal spin vectors can be written as
\begin{eqnarray}
   s_{1,2}^l   &=& \left[\beta,0,0,\pm 1 \right] /m_{\gw} \nonumber \\
   s_{1,2}^t   &=& \left[ 0,\pm 1,0,0 \right]             \nonumber \\
   s_{1,2}^n   &=& \left[ 0, 0,\pm 1,0 \right]                       \,.
\end{eqnarray}
The longitudinal $\pm$ components describe helicity + states of the
two gluinos; the spin vectors of the helicity $-$ states are given by
$- s^l_{1,2}$.  The matrix $\mathcal{C}$ is effectively $2 \times 2
\oplus 1 \times 1$ dimensional; it is symmetric for Majorana gluinos
and consists of four non-trivial components: $ll,lt=tl,tt;nn$.  Based
on the correlation tensor, before orthogonalization with respect to
the gluino momenta,
\begin{eqnarray}
   q\bar{q} \; {\rm channel}:\;\; {\mathcal{C}}_{\mu\nu} &=& \frac{\beta}{s} \left[ \frac{\alpha_s^2}{6}
                                                             \left(\beta^2\sin^2\theta g_{\mu\nu}-2 \kappa_{+} k_{1,\mu}k_{2\nu}
                                                             \right)\right.\nonumber\\
                                                         &+& \frac{\alpha_s \hat{\alpha}_s}{3}
                                                             \frac{2\beta^2\sin^2\theta g_{\mu\nu}-(m_{\gw}^2+2\kappa_-)
                                                              k_{2\mu}k_{1\nu}+(m_{\gw}^2-2\kappa_+)k_{1\mu}k_{2\nu}}
                                                              {m_{-}^2-2\kappa_+}\nonumber\\
                                                         &-& \left.\hat{\alpha}_{s}^2 \left(\frac{32}{27} \,
                                                             \frac{m_{\gw}^2 k_{2\mu}k_{1\nu}}{(m_{-}^2-2\kappa_{+})^2}
                                                             - \frac{4}{27} \, \frac{\beta^2\sin^2\theta g_{\mu\nu}
                                                             + (m_{\gw}^2-2\kappa_{+})k_{1\mu}k_{2\nu}}
                                                             {(m_{-}^2-2\kappa_{-})(m_{-}^2-2\kappa_{+})}\right) 
                                                              \right]^{\mathcal{S}} /{\mathcal{N}}_{q \bar{q}} \nonumber \\[2mm]
   gg       \; {\rm channel}:\;\; {\mathcal{C}}_{\mu\nu} &=& \frac{\alpha_{s}^2 \, \beta}{s} \, \frac{9}{32} \,
                                                             \left[\frac{1}{2} \left(- \kappa_{+}\kappa_{-} g_{\mu\nu}
                                                             +2\kappa_{+} k_{1\mu}k_{2\nu}\right)\right.\nonumber\\
                                                         &+& \frac{2m_{\gw}^2}{\kappa_{+}^2} \,\left((m_{\gw}^2-\kappa_{+})g_{\mu\nu}
                                                             +k_{2\mu}k_{1\nu}\right)                                                             
                                                             +\frac{1}{\kappa_{+} \kappa_{-}}\,
                                                             \left((m_{\gw}^4+\beta^2\sin^2\theta)g_{\mu\nu}
                                                             +(m_{\gw}^2-2\kappa_{+}) k_{1\mu}k_{2\nu}\right)\nonumber\\
                                                         &+& \left.\frac{1}{2\kappa_{+}} \, \left(2(\kappa_{+} \kappa_{-}
                                                              +m_{\gw}^2 \beta\cos\theta)g_{\mu\nu}+(m_{\gw}^2-2\kappa_{+})k_{1\mu}k_{2\nu}
                                                              -(m_{\gw}^2+2\kappa_{-})k_{2\mu}k_{1\nu}\right)
                                                              \right]^{\mathcal{S}} /{\mathcal{N}}_{gg}
\end{eqnarray}
with $[F]_{\mu \nu}^{\mathcal{S}}$ denoting the symmetrized tensor
\begin{equation}
[F]_{\mu \nu}^{\mathcal{S}}=F_{\mu\nu}(\cos\theta,k_1,k_2)+F_{\mu\nu}(-\cos\theta,k_2,k_1)\,,
\end{equation}
and $k_1,k_2$ being the initial parton 4-momenta, the spin
matrix-elements can easily be derived:
\begin{eqnarray}
q\bar{q} \; {\rm channel}:\;\;{\mathcal{C}}^{ll} &=& -\frac{\beta}{s}
                                                  \left[\frac{\alpha_{s}^2}{6} 
                                                  \left(\beta^2+(1+m_{\gw}^2)\cos^2\theta\right)
                                                  +\frac{2\,\alpha_{s} \hat{\alpha}_{s}}{3} \, 
                                                  \frac{\beta^2+2\beta\cos\theta+(1+m_{\gw}^2)\cos^2\theta}
                                                  {m_{-}^2-2\kappa_{+}}\right.\nonumber\\ 
                                              &&  \hspace{0.5cm} \left.+\hat{\alpha}_{s}^2 \left( \frac{32}{27} \, 
                                                  \frac{(\beta+\cos\theta)^2}{(m_{-}^2-2\kappa_{+})^2} 
                                                  +\frac{4}{27} \, \frac{m_{\gw}^2 \cos^2\theta}
                                                  {(m_{-}^2-2\kappa_{-})(m_{-}^2-2\kappa_{+})}\right)\right]^{\mathcal{S}}
                                                  /{\mathcal{N}}_{q\bar{q}} \nonumber\\[2mm]
   {\phantom{q\bar{q} \; {\rm channel}:\;\;}}
                           {\mathcal{C}}^{lt} &=& \frac{\beta m_{\gw}}{s} \sin\theta
                                                  \left[\frac{\alpha_s^2}{3} \cos\theta
                                                  + \frac{2\, \alpha_s \hat{\alpha}_s}{3} 
                                                  \frac{\beta+2\cos\theta}
                                                  {m_{-}^2-2\kappa_{+}}\right.\nonumber\\
                                              & & \left. \qquad + \hat{\alpha}_s^2 \left(
                                                  \frac{32}{27} \frac{\beta+\cos\theta}
                                                  {(m_{-}^2-2\kappa_{+})^2}+\frac{4}{27}\frac{\cos\theta}
                                                  {(m_{-}^2-2\kappa_{+})(m_{-}^2-2\kappa_{-})}
                                                  \right)
                                                  \right]^{\mathcal{A}} /{\mathcal{N}}_{q\bar{q}} \nonumber\\[2mm]
   {\phantom{q\bar{q} \; {\rm channel}:\;\;}}
                           {\mathcal{C}}^{tt} &=& -\frac{\beta}{s} \, \sin^2\theta
                                                         \left[ \frac{\alpha_{s}^2}{6} \, (m_{\gw}^2+1)
                                                         +\frac{2\alpha_{s} \hat{\alpha}_{s}}{3} \,
                                                         \frac{m_{\gw}^2+1}{m_{-}^2-2\kappa_{+}}\right.\nonumber\\
                                                     & & \hspace{1.7cm} +\left.\hat{\alpha}_{s}^2 \left( \frac{32}{27} \,
                                                         \frac{m_{\gw}^2}{(m_{-}^2-2\kappa_{+})^2}
                                                         +\frac{4}{27} \,\frac{1}{(m_{-}^2-2\kappa_{-})(m_{-}^2-2\kappa_{+})}
                                                         \right) \right]^{\mathcal{S}}
                                                         /{\mathcal{N}}_{q\bar{q}} \nonumber \\[2mm]
   {\phantom{q\bar{q} \; {\rm channel}:\;\;}}
                           {\mathcal{C}}^{nn} &=& \frac{\beta^3}{s} \sin^2\theta
                                                  \left[ \frac{\alpha_s^2}{6}  
                                                  + \frac{2\,\alpha_s \hat{\alpha}_{s}}{3} 
                                                  \frac{1}{m_{-}^2-2\kappa_+}
                                                  +\frac{4\,\hat{\alpha}_s^2}{27} 
                                                  \frac{1}{(m_-^2-2\kappa_+)(m_-^2-2\kappa_-)}
                                                  \right]^{\mathcal{S}}/{\mathcal{N}}_{q\bar{q}}
\end{eqnarray}                                                  
\begin{eqnarray}
   gg \; {\rm channel}:\;\; {\mathcal{C}}^{ll} &=& \frac{9 \alpha_s^2 \beta}{32 m_{\gw}^2 s}\,\left[
                                                   \kappa_+\,(\beta-\cos\theta)^2-\frac{\kappa_+ \kappa_-}{2}(1+\beta^2)
                                                   +\frac{2 m_{\gw}^2}{\kappa_+^2} \left((\beta+\cos\theta)^2
                                                   +(1+\beta^2)(m_{\gw}^2-\kappa_+)\right)\right.\nonumber\\
                                               & & \qquad \quad -\frac{1}{\kappa_+ \kappa_-}\left(
                                                   (2\kappa_--m_{\gw}^2)(\beta+\cos\theta)^2
                                                   -(1+\beta^2)(m_{\gw}^4+\beta^2\sin^2\theta)\right)\nonumber\\
                                               & & \qquad \quad +\frac{1}{2\kappa_+}\left(
                                                   (m_{\gw}^2-2\kappa_+)(\beta-\cos\theta)^2-(m_{\gw}^2+2\kappa_-)
                                                   (\beta+\cos\theta)^2\right.\nonumber\\
                                               & & \qquad\quad \left.\left.+ 2\ (1+\beta^2)
                                                   (\beta m_{\gw}^2\cos\theta+\kappa_+\kappa_-)\right)
                                                   \right]^{\mathcal{S}}/{\mathcal{N}}_{gg}\nonumber\\[2mm]
   {\phantom{gg \; {\rm channel}:\;\;}}
                           {\mathcal{C}}^{lt} &=& -\frac{9 \alpha_s^2 \beta}{32 s}\, \frac{\sin\theta}{m_{\gw}}\left[
                                                  \kappa_- \ (\beta+\cos\theta)
                                                  +\frac{2m_{\gw}^2}{\kappa_+^2} (\beta+\cos\theta)\right.\nonumber\\
                                              & & \qquad \qquad \quad \left.-\frac{1}{\kappa_+\kappa_-}
                                                  (\beta+\cos\theta)(2\kappa_--m_{\gw}^2)
                                                  -\frac{m_{\gw}^2}{\kappa_+} (\beta+2\cos\theta)
                                                  \right]^{\mathcal{A}}/{\mathcal{N}}_{gg}\nonumber\\[2mm]
   {\phantom{gg \; {\rm channel}:\;\;}}
                           {\mathcal{C}}^{tt} &=& \frac{9 \alpha_s^2 \beta}{32 s} \left[
                                                  \kappa_+\,(\sin^2\theta-\frac{\kappa_-}{2})
                                                  +\frac{2m_{\gw}^2}{\kappa_+^2}(m_{\gw}^2+\sin^2\theta-\kappa_+)
                                                  -\frac{1}{\kappa_+\kappa_-}
                                                  \left((2\kappa_+-1)\sin^2\theta-m_{\gw}^4\right)\right.\nonumber\\
                                              & & \left.\qquad \quad +\frac{1}{\kappa_+}
                                                  (\beta\,m_{\gw}^2\cos\theta+\kappa_+\kappa_--2\sin^2\theta)
                                                  \right]^{\mathcal{S}}/{\mathcal{N}}_{gg}\nonumber\\[2mm]  
   {\phantom{gg \; {\rm channel}:\;\;}}
                           {\mathcal{C}}^{nn} &=& -\frac{9 \alpha_s^2 \beta}{32 s} \left[
                                                  \frac{\kappa_+ \kappa_-}{2}
                                                  -\frac{2\,m_{\gw}^2 (m_{\gw}^2-\kappa_+)}{\kappa_+^2}
                                                  -\frac{m_{\gw}^4+\beta^2 \sin^2\theta}{\kappa_+\kappa_-}
                                                  -\frac{\kappa_{+}\kappa_{-}+m_{\gw}^2 \beta \cos\theta}{\kappa_+}
                                                  \right]^{\mathcal{S}}/{\mathcal{N}}_{gg}\,.
\end{eqnarray}
The index $l$ denotes positive gluino helicity.  The four correlation
matrix elements are displayed for ${\sqrt{{s}}} = 2$~TeV and SPS1a$'$
masses $M_{\gw}=607$~GeV and
$M_{\qw}=M_{\tilde{u}_L}=565$~GeV~\cite{SPS} in
Figs.~\ref{fig:qqcorr}(a) and (b) in the $q \bar{q}$ and $gg$
channels, respectively.  The antisymmetric terms drop out if
$q\qb\oplus\qb q$ channels are summed up for the symmetric $pp$
kinematics.  As expected, ${\mathcal{C}}^{ll}$, corresponding to $S_z
= 0$ for two equal gluino helicities in $q \bar{q} \to \gw\gw$,
approaches $-1$ for forward/backward production, lifted [in particular
by means of the $t/u$-exchange $\tilde{u},\tilde{d}$ propagators] to
larger values in between for non-zero gluino masses.
${\mathcal{C}}^{tt}$, on the other hand, is maximal for perpendicular
production.  The correlation of the normal polarizations is negative
for non-zero production angles but, as in the transverse direction,
approaches zero for forward production.  The $S_z = 0$ component of
the gluon-gluon spin wave-function in $gg \to \gw\gw$ gives rise to
maximal forward production of $S_z = 0$ gluino pairs and, as expected,
${\mathcal{C}}^{tt}$ and ${\mathcal{C}}^{nn}$ approach equal values
for forward production.

\subsection{Gluino Decays}

\noindent
If gluinos are heavier than squarks, they decay to quarks and squarks
\cite{BHZ}.  The squarks decay subsequently to quarks plus neutralinos
or charginos. Particularly at the reference point SPS1a/a$'$ the
lightest neutralino ${\tilde{\chi}}^0_1$ is almost a pure bino state,
while ${\tilde{\chi}}^0_2$ is almost a pure wino state, like the
chargino ${\tilde{\chi}}^\pm_1$. The heavier neutralinos and charginos
are higgsino-like states, and they do not play a significant role for
matter particles of the first two generations.  R-squarks will
therefore decay predominantly into the lightest neutralino
$\tilde{\chi}^0_1$ as LSP, while L-sqarks decay preferentially to the
heavier neutralino $\tilde{\chi}^0_2$ or chargino
$\tilde{\chi}^\pm_1$, followed by subsequent $\tilde{\chi}$ cascades.
The decay branching ratios \cite{sdecay} are collected in the
following set for the reference point SPS1a$'$:
\begin{equation}
\begin{array}{lclrclc}
   \Gamma[\gw \to q \qw_L] & = & 7.73\%\quad & \Gamma[\gw \to q \qw_R]
   & = & 17.0\%\quad & q \neq b,t\\
   \Gamma[\gw \to b {\tilde{b}}_1] & = & 10.8\% & \Gamma[\gw \to b
   {\tilde{b}}_2]\; & = & 4.67\% & \\
   \Gamma[\gw \to t {\tilde{t}}_1] & = & 9.81\%\,. & & & & 
\end{array}
\end{equation}
The decay $\gw \to t {\tilde{t}}_2$ is kinematically forbidden.

If the gluinos are polarized along the axis $\vec{\check{s}}$ in the
rest frame, final-state distributions are determined by the
polarization vector ${\mathcal{P}}_\mu$,
\begin{equation}
   d\Gamma = d\Gamma^{\rm unpol}\, [1- {\mathcal{P}}_\mu s_\mu ] \,,
\end{equation}
in a general Lorentz frame $\check{s}_\mu \to s_\mu$.

For gluinos polarized with degree unity, the angular distribution of
the quark jets with respect to the spin axis depends on the
particle/antiparticle character of the squarks, for the first two
generations:
\begin{eqnarray}
   \frac{1}{\Gamma}\, \frac{d\Gamma}{d\cos\theta} [\gw \to q_{R,L} \, \qw^\ast_{R,L}]  &=&
           \frac{1}{2}\, \left[ 1 \pm \cos\theta \right]                         \nonumber \\
   \frac{1}{\Gamma}\, \frac{d\Gamma}{d\cos\theta} [\gw \to \bar{q}_{L,R} \, \qw_{R,L}] &=&
           \frac{1}{2}\, \left[ 1 \mp \cos\theta \right]                         \,.
\label{eq:costheta}
\end{eqnarray}
The mass eigenstates of sparticles of the third generation, stop
particles in particular, are mixtures of R,L current eigenstates
\cite{BHPZ} . Denoting the ${\tilde{t}}_R$ component of the [light]
${\tilde{t}}_1$ wave-function by $\cos\theta_{\tilde{t}}$, the
coefficients of the $\cos\theta$-terms in Eqs.(\ref{eq:costheta}) are
altered from unity to
\begin{eqnarray}
   \alpha_{\tilde{t}_{1,2}} &=& \pm \cos 2\theta_{\tilde{t}} \, \beta_t \, \kappa^{-1}_{\tilde{t}_{1,2}}\,,  \\
\mbox{where}\quad   \kappa_{\tilde{t}_{1,2}} &=& 1 \pm 2 \sin 2\theta_{\tilde{t}} \, (1-\beta^2_t)^{1/2} \nonumber\,,
\end{eqnarray}
and $\beta_t$ is the velocity of the top quark in the gluino rest
frame. As expected, non-zero top mass and mixing dilute the
polarization effects.

As discussed previously, the spin information is washed out in
inclusive analyses of the final states.  Adding up quarks/antisquarks
and antiquarks/squarks, the $\cos\theta$-dependent terms cancel each
other. These final states can be discriminated however by tagging
either the charge of the quark/antiquark $t,b$ in the third
generation, or by tagging the squark/antisquark by measuring the
charges in chargino decays.  This method can successfully be applied
within the first [second] generation only if up- and down-states can
be distinguished since the states ${\tilde{u}}_L$ and
${\tilde{d}}^\ast_L$ generate final states with the same charge
topology but opposite quark/antiquark helicities. If, however, their
decay branching ratios are different, the spin-dependent contributions
of the two channels do not add up to zero anymore but come with the
spin analysis power $\kappa =
|BR_{\tilde{d}}-BR_{\tilde{u}}|/[BR_{\tilde{d}}+BR_{\tilde{d}}]\neq
0$. Though the partial widths for decays to charginos are the same by
isospin invariance, the total widths of $\tilde{u}$ and $\tilde{d} /
\tilde{d}^{\ast}$ may be different, nevertheless, for
${\tilde{\chi}}^0_2$ and ${\tilde{\chi}}^0_1$ not being pure wino and
bino states \cite{yuk}.

\subsection{Bypass: The Dirac Alternative}

\noindent
If the gluinos were Dirac particles \cite{CDFZ} as formulated in N=2
hybrid models, they could only be produced in the fermion-antifermion
mode of gluino-pair production:
\begin{eqnarray}
   q \bar{q},\; gg &\to& \gw_D \gw^c_D  \,.       
\end{eqnarray}
The decay channels would be restricted, by conservation of the Dirac charge, to
\begin{eqnarray}
   {\gw}_D   &\to& \bar{q}_L \qw_R \;\;\; {\rm and} \;\;\; q_L {\tilde{q}}^\ast_L    \\
   {\gw}^c_D &\to& \bar{q}_R \qw_L \;\;\; {\rm and} \;\;\; q_R {\tilde{q}}^\ast_R    \,.
\end{eqnarray}
Assuming, as done commonly, the $\tilde{q}_R, {\tilde{q}}^\ast_R$
squarks to decay into the lightest, invisible neutralino
${\tilde{\chi}}^0_1 =$~LSP, and tagging the $\tilde{q}_L,
{\tilde{q}}^\ast_L$ squarks, the spins of the {\it individual} gluinos
$\gw$ and $\gw^c$ can be reconstructed with spin-analysis power 1 in
the Dirac theory.  Unlike the Majorana theory, quark and antiquark
always come with opposite helicities in pairs.

\subsection{Spin-Phenomenology of Gluino-Pair Production}

\noindent
The spin-correlation effects in gluino-pair production will be
illustrated by the analysis of jet-jet invariant masses as a simple
indicator. Since spin correlations will play only a r\^{o}le in
precision measurements at LHC, our illustration is designed to be of
qualitative theoretical nature, without cuts on observables and QCD
radiative corrections, {\it etc.} Thus, only a coarse picture of spin
effects will be presented for illustration as [semi]realistic
experimental simulations are far beyond the scope of this theoretical
study.

Quite generally, spin correlation effects are described in the process
$pp \to \gw\gw \to final~state$ by the production correlation matrix
$\mathcal{C}$ and the two decay polarization vectors ${\mathcal{P}}_1$
and ${\mathcal{P}}_2$ \cite{kuhn}:
\begin{equation}
   d\sigma = d\sigma^{\rm unpol}\, [1 + {\mathcal{C}}_{\mu\nu} 
                                        {\mathcal{P}}_{1\mu} {\mathcal{P}}_{2\nu}] \,.
\end{equation}
The spin correlation affects in principle all final-state observables.

\begin{figure}[t]
\epsfig{figure=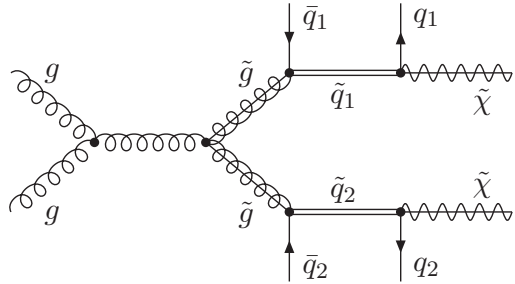, height=4cm}
\caption{Squark decay chain in gluino-pair production; chains
  involving anti-squarks are to be added.}
\label{fig:glugluchain}
\end{figure}

The decay chains of the two gluinos, cf. Fig.~\ref{fig:glugluchain},
\begin{eqnarray}
   \gw_1 \gw_2 &\to& [\bar{q}_1 \tilde{q}_1]\,[\bar{q}_2 \tilde{q}_2]
                \to   \bar{q}_1 {q}_1 \bar{q}_2 {q}_2 \,\tilde\chi\tilde\chi 
\end{eqnarray}
give rise to six invariant masses $M^2_{ij} = (p_i + p_j)^2$ which can
be formed within the four-jet ensemble of the final state.

The invariant mass distribution of the two near-jets, labeled
generically by $\bar{q}_1$ and $\bar{q}_2$, is most sensitive to the
gluino polarization. While it will not be known in practice which of
the observed final-state jets are associated with the near-jets in the
gluino decays, in the vast majority of events for SPS1a/a$'$-type mass
configurations these two jets can be identified with the jets of
minimal transverse momentum. Ordering therefore the jets $j_1$ to
$j_4$ according to rising transverse momenta, the $j_1 j_2$
combination is expected to retain most of the sensitivity to spin
correlations. Note that for gluino Majorana theories the $\bar{q}_1
\bar{q}_2$ combination comes in all possible configurations:
$\bar{q}_R \tilde{q}_L$ with $\bar{q}_R \tilde{q}_L $ and ${q_L}
\tilde{q}^\ast_L$, {\it i.e.} equal-helicity as well as
opposite-helicity (anti)quark states generating the low-$p_\perp$
jets. In contrast, Dirac theories would only allow opposite-helicity
(anti)quark states generating these jets.

To illustrate the effect of spin correlations we compare the jet-jet
invariant mass distributions for $\tilde{u}_L \tilde{u}_L$ and
$\tilde{u}_L \tilde{u}^\ast_L$ intermediate states, associated with
$\bar{q}_R \bar{q}_R$ and $\bar{q}_R {q}_L$ near-quark jets.  The L
squarks $\tilde{u}_L, \tilde{u}^\ast_L$ can be tagged by observing
leptonic decays of $\tilde{\chi}^\pm_1$ and $\tilde{\chi}^0_2$, which
discriminate L squarks from R squarks decaying to the invisible
$\tilde{\chi}^0_1$, cf. Ref.~\cite{CDFZ}. By tagging the L squarks,
kinematical effects due to different L/R squark masses, with size
similar to the spin effects, are eliminated.  The spin-correlations
will manifest themselves in different values of the jet-jet invariant
masses, which depend on the relative orientation of the gluino spins.
The average values $\langle M^2 \rangle = (\langle M^2_{\tilde{u}_L
  \tilde{u}_L} \rangle + \langle M^2_{\tilde{u}_L \tilde{u}^\ast_L}
\rangle)/2$ and the differences $\Delta M^2 = |\langle
M^2_{\tilde{u}_L \tilde{u}_L} \rangle - \langle M^2_{\tilde{u}_L
  \tilde{u}^\ast_L} \rangle|$ [the indices characterizing the
intermediate squarks] are presented in Table~\ref{tbl:gluglu} for all
six jet-jet invariant masses. The numerical results have been obtained
for the SPS1a$'$ scenario with masses $M_{\gw}=607$~GeV,
$M_{\tilde{u}_L}=565$~GeV, and $M_{\tilde\chi} =
M_{\tilde{\chi}_1^0}=98$~GeV. The CTEQ6L1 LO parton
densities~\cite{Pumplin:2002vw} have been adopted with the
corresponding leading-order $\alpha_s$, and all scales have been set
to $\mu = M_{\gw}$. All numerical results presented in this section
and below have been compared with results obtained using {\tt
  MadGraph/MadEvent}~\cite{MAD}; the results do agree with each other.

In the upper section of Table~\ref{tbl:gluglu} invariant masses for
identified jets are shown, and in the lower section for jets ordered
according to transverse momenta. For the SPS1a/a$'$-type mass
configurations considered here, the invariant mass distributions
involving near jets from the $\gw \to q\tilde{q}$ decays are
significantly softer than those involving far jets from $\tilde{q} \to
q\tilde{\chi}$ decays.  The gluino polarization affects the invariant
mass distribution involving the near jets, with a relative difference
between $\tilde{u}_L \tilde{u}_L$ and $\tilde{u}_L \tilde{u}^\ast_L$
intermediate states of about 10\%. For all other invariant mass
combinations the polarization effects are negligible. As evident from
the lower section of the table, the average invariant mass and the
invariant mass difference for the two jets with the smallest
transverse momentum $j_1j_2$ are very close to the corresponding
$\bar{q}_1 \bar{q}_2$ jet values, in concordance with general
expectations derived from the kinematics associated with $\qw,\gw$
mass parameters chosen in this example.

\begin{figure}
\epsfig{figure=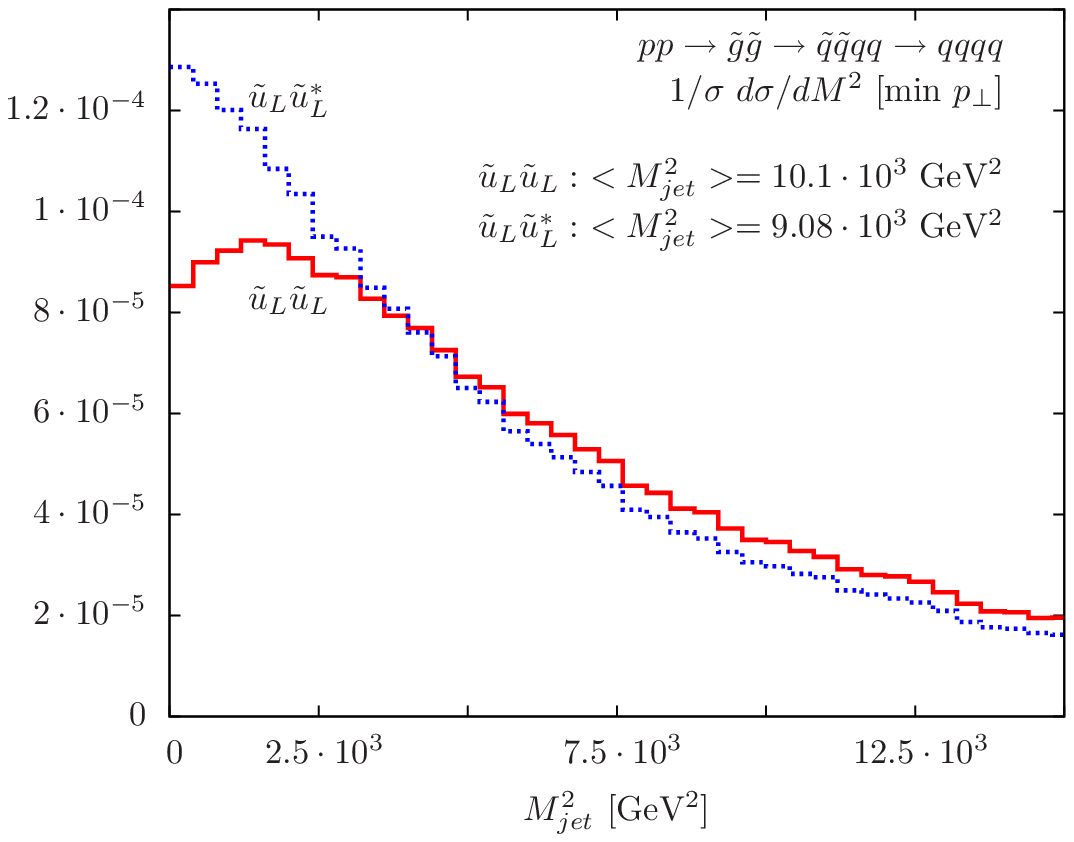, height=6.6cm}
\epsfig{figure=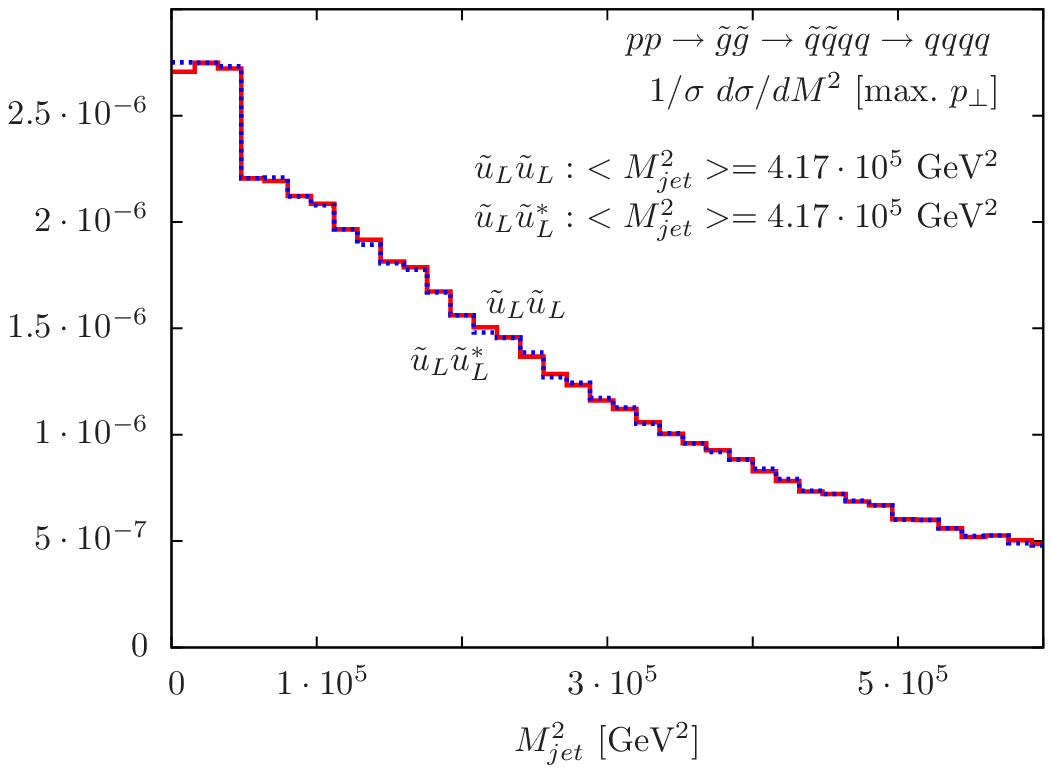, height=6.6cm}
\\[1ex]
\makebox[8.5cm]{(a)}
\makebox[7.5cm]{(b)}
\caption{Mass distribution of (a) the two jets with the lowest
  $p_\perp$, for which maximal spin correlations are predicted, and
  (b) the two jets with the highest $p_\perp$ where spin effects,
  within quark pairs generated almost exclusively by scalar decays,
  are expected to be very small.}
\label{fig:diffpT}
\end{figure}

The differential distribution of the $p_\perp$-ordered jet-jet
invariant mass $M^2_{j_1j_2}$ is depicted in Fig.~\ref{fig:diffpT}(a)
for constructive spin-correlation and contrasted with destructive
correlation.  Correlations among jets generated almost exclusively in
scalar squark decays are tiny as evident from the $M^2_{j_3j_4}$
distributions shown in Fig.~\ref{fig:diffpT}(b).  Without working out
the details it should be noted that cuts on the minimal missing energy
could be used to eliminate standard QCD stray jets. Lower $p_\perp$
cuts of order 50 GeV suppress additional QCD brems-strahl jets emitted
in the supersymmetric parton process itself; estimates indicate that
the signal of the spin-correlations is reduced by some 50\% when the
cuts are applied.

\begin{table}[ht!]
\centering
\vskip 0.5cm
\begin{tabular}{|l||cccccc|}
\hline
\multicolumn{ 7}{|l|}{Gluino pairs:$\gw\gw\to[\bar{q}\qw]\,[\bar{q}\qw]\to\bar{q}_1 q_1\bar{q}_2 q_2\,\tilde{\chi}\tilde{\chi}$}\\
\hline\hline
 original quarks                         & $\bar{q}_1 q_1$  &  $\bar{q}_1 \bar{q}_2$   & $\bar{q}_1 q_2$ 
                                         & $q_1 \bar{q}_2$  & \quad $q_1 q_2$          & $\bar{q}_2 q_2$ \quad \\
\hline
$\langle M^2\rangle\;{\rm [10^3\ GeV^2]}$&  23.9               & 9.17              & 62.4
                                         &  62.2               & 423               & 23.9                     \\
$\Delta M^2/\langle M^2\rangle\ [\%] $   &  $<$0.1             & 10.8              & 0.8
                                         &  0.8                & $<$0.1            & $<$0.1                     \\
\hline\hline
$p_\perp$ ordered jets                   & $j_1 j_2$           & $j_1 j_3$         & $j_1 j_4$
                                         & $j_2 j_3$           & $j_2 j_4$         & $j_3 j_4$                   \\
\hline
$\langle M^2\rangle\;{\rm [10^3\ GeV^2]}$ & 9.56               & 30.1              & 43.7
                                          & 39.6               & 64.7              & 417                      \\
$\Delta M^2/\langle M^2\rangle\ [\%] $    &  10.2              & 2.4               & 1.7
                                          &   2.0              & 3.0               & $<$0.1                      \\
\hline
\end{tabular}
\caption{\label{tbl:gluglu}
  Invariant jet-jet masses for gluino-pair production and
  decay. The average values $\langle M^2 \rangle$ and the differences
  $\Delta M^2$ of the invariant mass distributions 
  for $\tilde{u}_L \tilde{u}_L$ and
  $\tilde{u}_L \tilde{u}^\ast_L$ intermediate states are shown for 
  identified jets (upper section) and for jets ordered according to
  transverse momenta (lower section).}
\end{table}

\section{Super-Compton Process}

\subsection{Parton Level}

\noindent
Single polarization in gluino pair-production has been proved in the
foregoing section to be strongly suppressed at the squark-mass level
$\sim |M^2_L - M^2_R|/[M^2_L + M^2_R]$ for the two light generations
since the production process becomes effectively parity-even in the
limit $M_L \to M_R$. However, if the squarks are produced as final
particles, the L/R character can be identified and the
parity-violation in the Yukawa vertex becomes effective.  For example,
R-squarks may decay into the invisible LSP while L-squarks can be
marked by chargino decays. This constellation is realized in the
super-Compton process Eq.(\ref{eq:qwgw}).

Symmetrizing $qg \oplus gq$ as relevant for parton collisions at the
symmetric $pp$ collider LHC, the spin-summed total cross section
\cite{dawson,Beenakker,Hollik:2008vm} may be written
\begin{eqnarray}
   \sigma[q g \to \qw_{R,L} \gw] &=& \frac{\pi \alpha_s \hat{\alpha}_s}{s} \left[
                                   \left(1+\frac{m_-^2}{2}-\frac{\mqw^2 m_-^2}{8}\right)L_+
                                   + \left(\frac{2 m_-^2}{9}-\frac{\mqw^2 m_-^2}{8}
                                   - \frac{m_-^4}{18}\right)L_- \right.\nonumber\\
                               & & \hspace{1.2cm}\left.-p \left(\frac{7}{9}+\frac{32 m_-^2}{9}\right) \right]
\end{eqnarray}
with 
\begin{equation}
L_\pm=\log\frac{1+p\pm m_-^2/4}{1-p\pm m_-^2/4} \,,
\end{equation}
while the angular dependence reads:
\begin{eqnarray} 
   \frac{d\sigma}{d\Omega} [q g \to \qw_{R,L} \gw] &=& \frac{\alpha_s \hat{\alpha}_s\, p}{s} \left[ 
                                                       \frac{1}{36} \, \kappa_{\gw_{-}}
                                                       -\frac{1}{36 \kappa_{\qw_{+}}^2} (m_{\qw}^2-\kappa_{\qw_{+}})
                                                       (m_{-}^2+2\kappa_{\qw_{+}})\right.\nonumber\\
                                                   & & \qquad \quad - \frac{1}{16 \, \kappa_{\gw_{-}}^2} 
                                                       \left[m_{\tilde{g}}^2 (m_{-}^2-2\kappa_{\gw_{-}})
                                                       -2\kappa_{\gw_{-}}(\kappa_{\gw_{-}}+\kappa_{\qw_{+}})\right]
                                                       \nonumber\\
                                                   & & \qquad \quad +\frac{1}{576 \, \kappa_{\qw_{+}}}
                                                       \left[m_{-}^4-2(\mgw^2+\mqw^2)+2\kappa_{q_+}
                                                       (m_{-}^2+2)\right]\nonumber\\
                                                   & & \qquad \quad +\frac{1}{64 \, \kappa_{\gw_{-}}} \left[
                                                       m_{\tilde{g}}^4-2m_{\tilde{g}}^2(m_{\qw}^2+\kappa_{\gw_{-}})
                                                       +(m_{\qw}^2-4)(m_{\qw}^2+2\kappa_{\gw_{-}})\right]\nonumber\\
                                                   & & \left. \qquad \quad + \frac{1}{32 \, \kappa_{\qw_{+}} \kappa_{\gw_{-}}}
                                                       \left[m_{\qw}^4-m_{\tilde{g}}^4+ \kappa_{\gw_{-}} (m_{\qw}^2+m_{\tilde{g}}^2)
                                                       -2 \kappa_{\qw_{+}}(\mqw^2+\kappa_{\gw_{-}})\right]\right]^{\mathcal{S}} \,,
\label{eq:angdisqg}
\end{eqnarray}
where the abbreviations
\begin{eqnarray}
   \kappa_{\qw_{\pm}} &=& \E_{\qw} (1 \pm \beta_{\qw}\cos\theta)                  
\end{eqnarray}
and
\begin{eqnarray}
m_{\qw}     &=& 2 M_{\qw} / \sqrt{s} \hspace{2.2cm} \E_{\qw} = 2 E_{\qw} / \sqrt{s} = 1+(m_{\qw}^2-\mgw^2)/4  \nonumber\\
\beta_{\qw} &=& p/\E_{\qw}           \hspace{3cm}   p = 
                                     \sqrt{\left[1-(m_{\qw}+\mgw)^2/4 \right] \left[1-(m_{\qw}-\mgw)^2/4\right]}
\end{eqnarray}
have been used; $\qw \Leftrightarrow \gw$ correspondingly.  These
spin-averaged cross sections, discussed in detail in
Ref.~\cite{Beenakker}, are form-identical for $\qw_L$ and $\qw_R$
production.

\begin{figure}[t]
\epsfig{figure=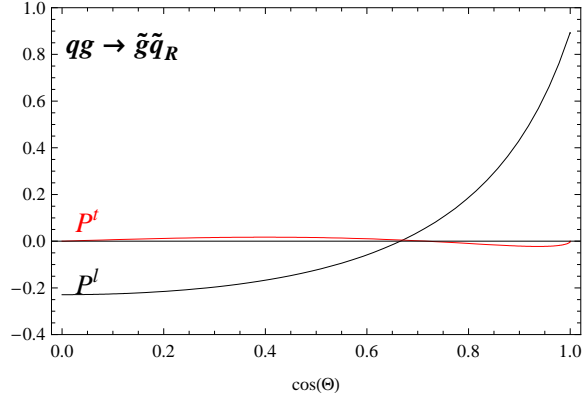, height=5.3cm}
\caption{Longitudinal and transverse components of the gluino polarization vector
         for R-squarks in the final state of the super-Compton process.}
\label{fig:qgcorrelation}
\end{figure}

The polarization vector of the gluino,
\begin{equation}
   \frac{d\sigma(s)}{d\Omega} =\frac{d\sigma}{d\Omega}\,\frac{1}{2} \,
                       \left[1 - {\mathcal{C}}_{\mu} s^{\mu} \right]    \,,
\end{equation}
using the spin-vector notation within the frame introduced before, can
easily be determined:
\begin{eqnarray}
   {\mathcal{C}}_{\mu} &=& \frac{\alpha_s \hat{\alpha}_s\,p\,\mgw}{s} \left[ 
                           -\frac{1}{36} k_{1\mu}
                           +\frac{1}{8 \, \kappa_{\gw_{-}}^2} \left(
                           (\kappa_{\gw_{+}}-2) k_{1\mu}
                           +(\mgw^2-\kappa_{\gw_{-}})k_{2\mu}\right) \right.\nonumber\\
                       & & \hspace{1.85cm} + \frac{1}{72 \, \kappa_{\qw_{+}}^2}
                           \left[\mgw^2+\mqw^2-2(p^2-\E_{\qw}^2+2(1+p\cos\theta))\right]k_{2\mu}\nonumber\\
                       & & \hspace{1.85cm}- \frac{1}{32 \, \kappa_{\gw_{-}}} \left[
                           (\kappa_{\gw_{+}}-4) k_{1\mu} 
                           +(\kappa_{\gw_{+}}-2\E_{\qw})k_{2\mu}\right]\nonumber\\
                       & & \hspace{1.85cm} - \frac{1}{576 \, \kappa_{\qw_{+}}} \left[
                           2 (2+p\cos\theta-\E_{\qw}) k_{1\mu}
                           + 2 (4-3\E_{\qw}+p\cos\theta) k_{2\mu}\right]\nonumber\\
                       & & \left. \hspace{1.85cm} + \frac{1}{32 \, \kappa_{\gw_{-}} \kappa_{\qw_{+}}}
                           \left[ 2 (2+p\cos\theta-\E_{\qw}) k_{1\mu} 
                           +2(\mgw^2-p^2+\E_{\qw}^2)k_{2\mu}\right]\right]^{\mathcal{S}}
                           /{\mathcal{N}}_{qg}  \,.
\end{eqnarray}
${\mathcal{N}}_{qg}$ denotes the symmetrized angular distribution
Eq.(\ref{eq:angdisqg}).  Evaluation of the vector in the gluino rest
frame yields the final results for the longitudinal, transverse and
normal components:
\begin{eqnarray}
   {\mathcal{P}}^l &=& \frac{\alpha_s \hat{\alpha}_s\, p}{s} \left[
                       \frac{1}{36} (\E_{\gw}\cos\theta-p)
                       +\frac{1}{8 \, \kappa_{\gw_{-}}^2} \left[
                       (2\E_{\gw}-\E_{\qw}\,\mgw^2)\cos\theta 
                       +p\,(\mgw^2-2)\right] \right.\nonumber\\
                   & & \qquad \quad -\frac{1}{72 \, \kappa_{\qw_{+}}^2}\left[
                       (p+\E_{\gw}\cos\theta)(3\,(2 \E_{\gw}-\mgw^2)
                       -(2\E_{\qw}+\mqw^2)+4p\cos\theta)\right]\nonumber\\
                   & & \qquad \quad - \frac{1}{16\,\kappa_{\gw_{-}}}\left[
                       (p^2+\E_{\gw}^2)\cos\theta-2p\E_{\qw}\right]
                       -\frac{1}{288 \, \kappa_{\qw_{+}}} \left[ 
                       2(p^2+\E_{\gw}(1-\E_{\qw}))\cos\theta+2p\,(3-2\E_{\qw})\right]\nonumber\\
                   & & \left. \qquad \quad - \frac{1}{16 \,\kappa_{\qw_{+}}\kappa_{\gw_{-}}}
                       [\E_{\gw}\,p\cos^2\theta-\frac{p}{4} (5\mgw^2+3\mqw^2+4)
                       +\frac{1}{4}\ (\mqw^4-\mgw^4-4\mqw^2)\cos\theta]\right]^{\mathcal{S}}
                       /{\mathcal{N}}_{qg}\nonumber\\[2mm]
   {\mathcal{P}}^t &=& \frac{\alpha_s \hat{\alpha}_s\,p\,\mgw\,\sin\theta }{s}\left[
                       \frac{1}{8\,\kappa_{\gw_{-}}^2}\left[2(\E_{\gw}-1)-\mgw^2\right]                       
                       -\frac{1}{72\,\kappa_{\qw_{+}}^2} \left[
                       3(\mgw^2-2\E_{\gw})+(\mqw^2+2\E_{\qw})-4p\cos\theta\right]
                       \right.\nonumber\\
                   & & \left.\hspace{2.3cm}+\frac{\E_{\gw}}{16\,\kappa_{\gw_{-}}}
                       +\frac{\E_{\gw}-\E_{\qw}}{288\,\kappa_{\qw_{+}}} 
                       - \frac{1}{16\,\kappa_{\qw_{+}}\kappa_{\gw_{-}}}
                       \left[2(\mgw^2-2\E_{\gw})+\E_{\qw}+2-p\cos\theta\right]\right]^{\mathcal{A}}
                       /{\mathcal{N}}_{qg}\nonumber\\[2mm]
   {\mathcal{P}}^n &=& 0     
\end{eqnarray}
at the Born level for R-squarks in the final state, while the overall
signs flip for L-squarks.  The non-trivial longitudinal and transverse
components of the polarization vector are displayed in
Fig.~\ref{fig:qgcorrelation} for the parton invariant energy $\sqrt{s}
=$ 2 TeV and the SPS1a$'$ masses $M_{\gw}=607$~GeV and
$M_{\qw}=M_{\tilde{u}_R}=547$~GeV.  While the transverse polarization
is very small for all production angles, the maximal parity violation
of the Yukawa vertex renders the degree of longitudinal polarization
large in the forward direction.

Normal gluino polarization is zero in Born approximation, but it can
be generated by vertex loop-corrections which render the transition
matrix formally $T$-odd. For example, the triangular $\qw \gw g$
super-QCD vertex correction to the amplitude $g q \to \gw\qw$ is
complex, but the size of the normal polarization remains suppressed as
a higher-order effect \cite{kuhn} at a level of $\alpha_s \tilde{M} /
\sqrt{s}$, with $\tilde{M}$ denoting the typical supersymmetry mass
scale.

The Dirac theory would generate, in the restricted set of processes
$qg \to \qw_R \gw^c_D$ and $\qw_L \gw_D$, the same degree of
polarization.
 
The single polarization vector can be determined experimentally by
using the two methods discussed in the previous section:

\noindent {\bf (a)} Tagging of the ${\tilde{u}}_L, {\tilde{d}}^\ast_L$
or ${\tilde{d}}_L, {\tilde{u}}^\ast_L$ squarks by searching for
$\ell^+$ or $\ell^-$ final states allows to reconstruct the spin with
analysis power $\kappa =
|BR_{\tilde{d}}-BR_{\tilde{u}}|/[BR_{\tilde{d}}+BR_{\tilde{d}}]$ in
the first two generations.

\noindent {\bf (b)} Analysis of stop and sbottom final states allows
the reconstruction of the spin vector, without dilution by destructive
interference effects of different flavors but with some dilution due
to the superposition of near and far-top quarks.

\subsection{Spin-Phenomenology of the Super-Compton Process}

\begin{figure}[t]
\epsfig{figure=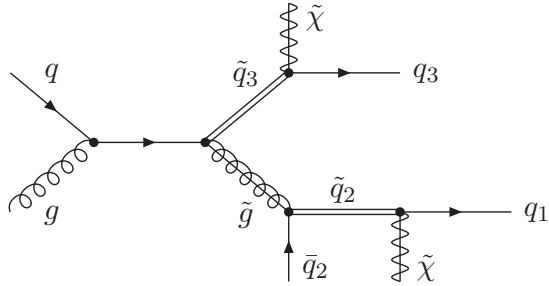, height=4cm}
\caption{Decay chains in the super-Compton process. While $q$ is
  effectively restricted to valence $u,d$ quarks, the squark $\qw_2$
  may be substituted by the corresponding anti-squark, the attached
  [anti]quarks correspondingly.}
\label{fig:scchain}
\end{figure} 

\noindent
The spin effects in the super-Compton process are described by the
polarization vectors ${\mathcal{C}}_\mu$ and ${\mathcal{P}}_\mu$ in
the gluino production and decay processes \cite{kuhn}:
\begin{equation}
   d\sigma = d\sigma^{\rm unpol} [1 - {\mathcal{C}}_\mu {\mathcal{P}}_\mu ]  \,.
\end{equation}
Depending on the sign of the product of the production and decay
polarization vectors either constructive or destructive spin effects
are generated, affecting, in principle, all experimental observables.

In parallel to the foregoing gluino-pair discussion we illustrate the
spin phenomenology in the super-Compton process again by analyzing
jet-jet invariant masses. The reference jet will be the primary squark
decay jet $q_3$ recoiling against the gluino decay antiquark jet
$\bar{q}_2$ and the secondary quark decay jet $q_1$ emitted in the
squark decay of the gluino chain, cf. Fig.~\ref{fig:scchain}:
\begin{equation}
   \qw\gw \to [\qw_3]\,[\bar{q}_2 \qw_2] \to q_3 \bar{q}_2 q_1 \tilde\chi \tilde\chi   \,. 
\end{equation}
The incoming quarks will be taken as $u,d$ valence quarks.  Three
invariant masses $q_ 1\bar{q}_2, q_1 q_3, \bar{q}_2 q_3$ can be formed
from the three final-state quark momenta.  Since the polarization
vector $\mathcal{P}$ can be varied by picking L squarks or antisquarks
in the $\gw$ decay state, different values are predicted for the
jet-jet invariant masses.

Gluino Majorana theories generate equal- and opposite-helicity
(anti)quark final states $\bar{q}_2 q_3$ while Dirac theories restrict
these final states to equal-helicity pairs.

The average values of the three jet-jet invariant masses for
$\tilde{u}_L \tilde{u}_L$ and $\tilde{u}_L \tilde{u}^\ast_L$ final
states coming with $u_L$ and $u_L/{\bar{u}_R}$ quark jets are
presented in Tab.\ref{tbl:SuperC}, for identified jets in the upper
row, and jet pairs ordered according to rising invariant masses
$M_{\rm inv}$ in the second row.

\begin{table}[ht!]
\centering
\vskip 0.5cm
\begin{tabular}{|l||ccc|}
\hline
\multicolumn{4}{|l|}{Super-Compton:$\qw\gw\to[\qw]\,[\bar{q}\qw]\to q_3\bar{q}_2 q_1 \,\tilde{\chi}\tilde{\chi}$}\\
\hline\hline
 original quarks & $q_1 \bar{q}_2$ & $q_1 q_3$ & $ \bar{q}_2 q_3$ \\
\hline
$\langle M^2 \rangle \; {\rm [10^4\ GeV^2]}$ & 2.39   & 48.3  & 7.13 \\
$\Delta M^2/\langle M^2\rangle \ [\%]$ & $<$0.1 &  0.8  & 7.9   \\
\hline\hline
$\rm M_{\rm inv}$ ordered jets &\; small\; &\; medium\; &\; large\; \\
\hline
$\langle M^2 \rangle \; {\rm [10^4\ GeV^2]}$ & 1.80 & 7.33 & 48.7 \\
$\Delta M^2/\langle M^2\rangle\ [\%] $ & 5.0   & 5.9   & 0.7 \\
\hline\hline
$\rm{M_{jjj} > 2.5}\ \rm{TeV:}$ & small & medium & large \\
\hline
$\langle M^2 \rangle \; {\rm [10^4\ GeV^2]}$ & 2.46 & 76.3 & 775 \\
$\Delta M^2/\langle M^2\rangle\ [\%] $ & 10.3   & 21.2 & 2.0 \\
\hline
\hline
\end{tabular}
\caption{\label{tbl:SuperC}
  Invariant jet-jet masses in the super-Compton process. 
  The average values $\langle M^2 \rangle$ and the differences
  $\Delta M^2$ of the invariant mass distributions 
  for $\tilde{u}_L \tilde{u}_L$ and
  $\tilde{u}_L \tilde{u}^\ast_L$ intermediate states are shown for 
  identified jets (upper section) and for jets ordered according to
  invariant mass (middle section). The lower section shows the 
  enhanced polarization effects if large invariant masses 
  $M_{jjj} > 2.5$~TeV 
  are selected.}
\end{table}

\begin{figure}
\epsfig{figure=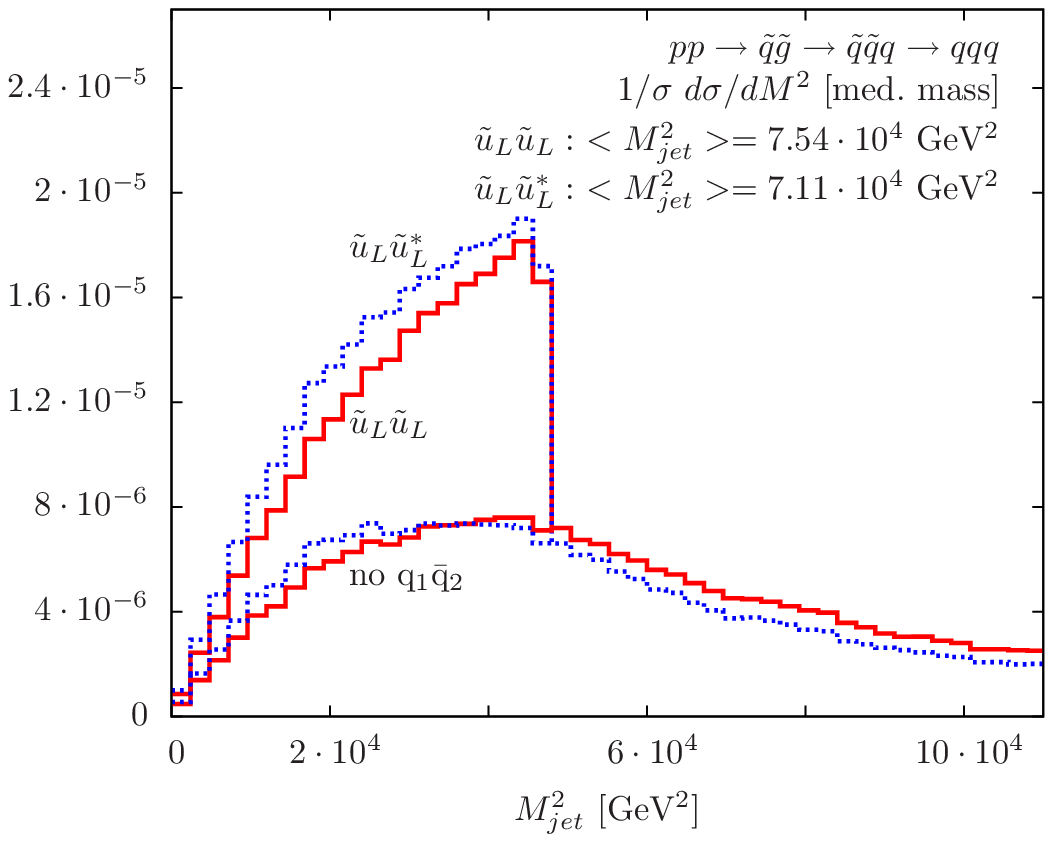, height=6.6cm}
\epsfig{figure=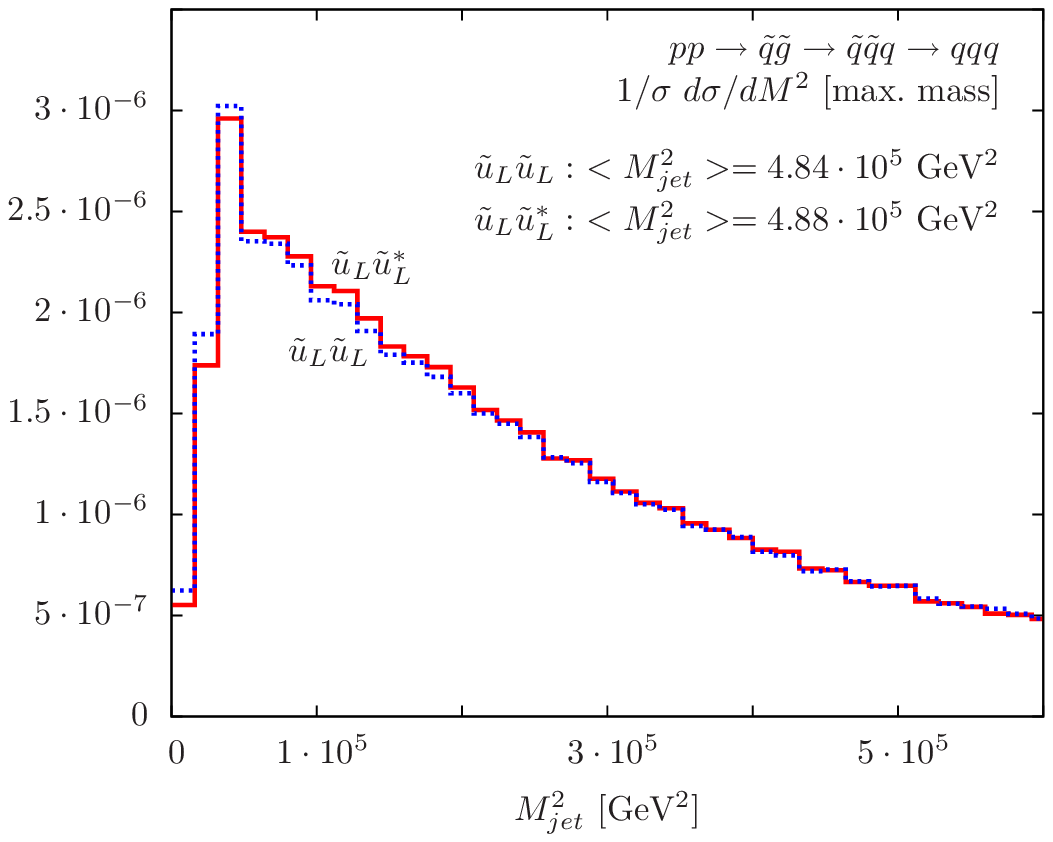, height=6.6cm}
\\[1ex]
\makebox[9cm]{(a)}
\makebox[7.5cm]{(b)}
\caption{Mass distribution of (a) the second largest invariant mass;
  the same-side $q_1 \bar{q}_2$ parton combination is subtracted in
  the lower curves; (b) the distribution for jets of maximal invariant
  mass, corresponding largely to partons $q_1 q_3$.}
\label{fig:diffmass}
\end{figure}

As evident from the table, the combination $\bar{q}_2 q_3$, involving
the primary antiquark decay jet of the polarized gluino, provides the
highest sensitivity to spin effects. This combination is mapped, on
the average, to the second largest invariant mass in the $M_{\rm inv}$
ordered three-jet ensemble. A clear distinction emerges between
constructive and destructive spin effects in this observable.  If
large invariant masses for the three-jet final states are selected,
say $M_{jjj} >$ 2.5 TeV, the longitudinal gluino polarization is
greatly enlarged, boosting forward or backward the recoiling decay
squark, and raising or lowering the jet-jet invariant masses
accordingly.  The spin-dependent distributions of the second largest
invariant mass are depicted in Fig.~\ref{fig:diffmass}(a).  [The
contamination due to same-side $\bar{q}_2 q_1$ partons, which generate
the wedge with the standard sharp edge at $\approx [M^2_{\gw} -
M^2_{\qw}]$ but do not give rise to spin asymmetries, is subtracted
for illustration in the two lower curves.]  Correlations among squark
decay jets $[q_1 q_3]$ are small, Fig.~\ref{fig:diffmass}(b), as
expected.

\section{SUMMARY}

\noindent
Spin correlations in gluino-pair production and polarization effects
in single gluino production of the super-Compton process affect the
distributions of the experimentally observed final states. In this
report we have analyzed the theoretical basis of these effects,
calculating the two-gluino spin-correlation matrix in the first case
and the gluino polarization vector in the second. A few examples for
jet invariant masses illustrate the size of these effects at the
theoretical level. They become relevant only if L and R squarks,
coming in association with R/L polarized antiquarks, {\it etc.}, are
discriminated by measuring, for instance, charges in the third
generation.  The spin effects are modest, typically of about 10\% in
the spin-sensitive observables. Nevertheless, when the LHC potential
is fully exploited for precision measurements and analyses are
performed to determine supersymmetry parameters like mixings,
couplings, {\it etc.}, such effects must be controlled properly in the
analyses of the measured final states.  The report is intended to
provide a first step in this direction.

%
\acknowledgments{}

\noindent
We acknowledge helpful discussions with G.~Polesello on experimental
aspects of this study.  PMZ is grateful to the Institut f\"ur
Theoretische Physik E for the warm hospitality extended to him at the
RWTH Aachen.

\vskip 0.3cm

%

\end{document}